\documentclass[aoas,MSNbibl,nameyear,dvips]{arximspdf}

\doi{10.1214/10-AOAS449}
\volume{5}
\issue{1}
\pubyear{2011}
\firstpage{1}
\lastpage{4}

\begin{document}
\begin{frontmatter}

\title{Editorial}
\runtitle{Editorial}

\begin{aug}
\author{\fnms{Michael L.} \snm{Stein}\corref{}\ead[label=e1]{stein@galton.uchicago.edu}}
\runauthor{M. L. Stein}
\affiliation{University of Chicago}
\address{Department of Statistics\\
University of Chicago\\
Chicago, Illinois 60637\\
USA\\
\printead{e1}} 
\end{aug}

\received{\smonth{12} \syear{2010}}



\end{frontmatter}

Many of you reading these words will have been attracted by the discussion
paper [McShane and Wyner (\citeyear{mcshane})], in which case, this may be the first, but
hopefully not the last, time you will have read anything in a statistics
journal.  I would like to take this opportunity to discuss the review
process in our journal and to make some comments about the role of
statistics and uncertainty assessment in paleoclimatology and the broader
debate about climate change.

Most papers that are published in this journal, including McShane and Wyner
(\citeyear{mcshane}) are sent by an editor (in this case, me) to an Associate Editor who
then seeks the input of one or more referees.  The referees write reports
giving their opinion of the work, including recommendations for how it
might be improved if it were to be published.  Taking into account the
reports of the referees and his or her own reading of the paper, the
Associate Editor generally writes an additional report and makes a
recommendation to the editor as to the suitability of the paper for
publication in the journal.  In addition to synthesizing the opinions of
the Associate Editor and the referees, I look through the paper myself
and often add my own commentary to those of the other reviewers and make
a decision about the publication status of the paper.  Even when the
recommendation is favorable, we generally request revisions
and, in fact, during my term as an editor, I have not accepted a single
paper without asking for at least some changes.

When an editor accepts a paper, it does not mean that the journal or the
individual editor personally endorses or agrees with it.  Indeed, we commonly
publish papers that one or more of the reviewers, including the editor,
will disagree with in part.  Acceptance of a paper reflects our opinion
that the work represents a meaningful contribution to applied statistics,
broadly construed, and that the authors have made a good faith effort to
respond to the concerns of the reviewers.

McShane and Wyner (\citeyear{mcshane}) received a careful reading by two referees, an
Associate Editor and myself.  All four of us made detailed comments
about aspects of the paper that we wanted to see changed before we could
recommend publication in \textit{The Annals of Applied Statistics}.
The authors undertook an extensive
revision of their work and the paper was reviewed again by all of the
original reviewers as well as by Tilmann Gneiting, an incoming editor
at the journal and, after additional minor changes, I accepted the paper.

Because of the obvious interest in this paper's subject matter, I decided to
make it a discussion paper.  After consulting with some members of the
editorial board and a few others including the authors, I invited
a broad range of individuals with an interest in the topic to contribute
discussions.  All but one of the people I invited contributed a
discussion.  In the interest of moving the publication process along and
keeping the discussions focused, I gave discussants one month to write their
discussion and asked that they keep the text of their discussion
(excluding figures and references) to about two pages in length.  Many of
the discussants could and would have written more lengthy discussions
had I permitted it.  Unlike papers, discussions and the authors'
rejoinder do not
generally undergo a detailed review although, in this case, Tilmann
Gneiting and I did carefully read through this material and I asked for
some changes in presentation or occasional cuts to keep the discussions
close to the page limit I imposed.
In contrast,
the supplementary (online) material provided by the authors and the discussants
has not been meaningfully reviewed.  I would like to thank the discussants
for their impressive (by academic standards) willingness to conform to
my requests to keep their discussions short and to submit their discussions
on time.
I would also like to acknowledge the enormous effort by the authors to write
their detailed rejoinder in about two months.

Anyone who reads this paper and the ensuing discussion should realize
that there is more to be said about statistical methods for paleoclimate
reconstruction.  Some of this work will hopefully appear in this journal,
so stay tuned.  I would encourage those of you who want to work in this area
to focus on developing
new and better ways for carrying out climate reconstructions
using all of the available information rather than rehashing the merits
of previous approaches.  Several of the discussants make useful recommendations
in this regard.

Based on my experiences handling this paper and my other engagements
in climate change, I would like to make a few specific and, hopefully,
uncontroversial recommendations.

\begin{itemize}
\item Greater cooperation between the climatological and statistical
communities would benefit both disciplines and be invaluable in the broader
public discussion of climate change.  There have been great strides made
in this regard in recent years, which is reflected in the diversity of
affiliations of the discussants and the extent to which they demonstrate
their understanding of both statistics and climatology.  Hopefully the
present discussion paper will only help to spur further cooperation
between the disciplines.
\item There is a movement in various disciplines to make
all numerical results reported on in published papers reproducible by
providing all of the data and code used to generate the results [Diggle
and Zeger  (\citeyear{diggle}); Fomel and Claerbout  (\citeyear{fomel}); Peng  (\citeyear{peng})].
One could debate whether this reproducibility is desirable for all research,
but it should be a requirement for research that has potentially
important public policy implications whenever permissible (e.g., does not
violate privacy rules).  The authors and those discussants who
report numerical results
have provided their data and code, all of which is archived at
\href{http://www.imstat.org/aoas/supplements/default.htm}{http://www.imstat.org/aoas/}
\href{http://www.imstat.org/aoas/supplements/default.htm}{supplements/default.htm}.
However, in some circumstances it may be important to provide even further
information.  In particular, if the data used in a study were subject
to any preprocessing and/or selection from a larger database, it may be
critical to detail this process.
For analyses that entail processing through multiple programs and/or
packages, it may be, as the authors note in their rejoinder,
quite difficult to provide sufficient information
to make the work truly reproducible by a typical user.
In some circumstances, authors should report on statistical
analyses that were tried out but discarded for various reasons.  Of course,
it would be unreasonable and unhelpful to ask authors to document every
analysis they tried, but when the stakes are sufficiently high, authors could
document analyses that were seriously considered
but, for whatever reasons, deemed inferior to the published analysis.
\item As I get older, I find myself saying many of the same things in every
class I teach.  One claim I frequently make is that, in terms
of what is most important about using statistics to answer scientific
questions, data are more important than models and models are more important
than specific modes of inference.  In the present context, this suggests
focusing efforts on the development of new climate proxies and the attendant
statistical issues in processing them into usable forms.  More broadly,
statisticians need to engage the entire climatological
community in questions of what raw data to collect and in
how to process these data into forms that can be broadly used.
\end{itemize}

Some of the discussants touch on the broader implications of
paleoclimate reconstruction for the study of climate change. I would just
like to raise one further issue, again related to something I tend to say
in every class I teach: classical statistical hypothesis testing is
overused in the scientific literature. I particularly
object to the testing of sharp null hypotheses when there
is no plausible basis for believing the null is true.  An example of an
implausible sharp null hypothesis would be that a large increase in the
concentration of $\mbox{CO}_2$ in the atmosphere has \textit{exactly
zero} effect on
the global mean temperature.  When a null hypothesis of no effect
is untenable, emphasis should be on
estimation and/or prediction along with uncertainty quantification.
Thus, the testing and attribution questions
for climate change seem to me to be irrelevant and the focus needs to be
on prediction.
Seen in this light, paleoclimate reconstructions on a range of time scales
are more useful for estimating the effect of various climate
forcings (e.g., solar variability, aerosols and trace gases) on
the climate than for testing sharp null hypotheses.
Appropriate assessment of uncertainties in reconstructions of both
the climate and the forcings are, of course, critical to this endeavor.

Statisticians are, by their professional nature, skeptics.  We often find
that researchers in other fields have not taken proper account of all
important elements of uncertainty when they analyze data.  However,
uncertainty is not a basis for inaction.  (If it were, none of us would get
out of bed in the morning.)  Taking appropriate account of the uncertainties
about the future climate,
we need to be evaluating the consequences of various courses of action
by making the best use of all of our knowledge about climatology and the
many other disciplines that bear on the issue.
Careful study of tiny pieces of the knowledge base is important, but
no single study provides a direct basis for action or inaction.
In particular, the presence of even substantial
uncertainties does not necessarily mean that the
appropriate response is to wait for better information about the future
climate.
Any potential benefits of waiting depend in part on estimates of how much our
uncertainty is likely to decrease over the next several years.
My understanding
is that the major uncertainties in climate projections on time scales
of more than a few decades are unlikely to be resolved in the near future.
Thus, while research on climate change should continue,
now is the time for individuals and governments to act to limit the
consequences of greenhouse gas emissions on the Earth's climate over
the next century and well beyond.


\printaddresses

\end{document}